\documentclass{article}
\usepackage{arxiv}
\usepackage{url}
\usepackage[bookmarks=false]{hyperref}
 \usepackage[final]{pdfpages}
\usepackage{comment}
  \usepackage{caption,subcaption}
  \usepackage{xcolor}
 \usepackage{graphicx}
\definecolor{Orange}{rgb}{1,0.5,0}
\definecolor{Red}{rgb}{1,0,0}
\definecolor{Green}{rgb}{0,0.65,0.5}
\definecolor{Purple}{rgb}{0.75,0,1}
\definecolor{babypink}{rgb}{0.96, 0.76, 0.76}
\definecolor{azure}{rgb}{0,0.49,1}
\definecolor{periwinkle}{rgb}{0.8, 0.8, 1.0}
\definecolor{Pink}{RGB}{255, 102, 204}
\newcommand{\add}[1]{\textcolor{black}{#1}}

\newcommand{\elissa}[1]{\textsf{\textbf{\textcolor{Purple}{[[EMR: #1]]}}}}

\AtBeginDocument{%
  \providecommand\BibTeX{{%
    \normalfont B\kern-0.5em{\scshape i\kern-0.25em b}\kern-0.8em\TeX}}}

\begin{document}

\title{Risk, Resilience and Reward:\\Impacts of Shifting to Digital Sex Work}


\author{ Vaughn Hamilton\\Max Planck Institute for Software Systems\\
\texttt{vhamilto@mpi-sws.org}\\
\And
Hanna Barakat\\
Brown University\\
\texttt{hanna\_barakat@brown.edu}\\
\And
Elissa M. Redmiles\\
Max Planck Institute for Software Systems\\
\texttt{eredmiles@mpi-sws.org}
}

\maketitle
\begin{abstract}
  Workers from a variety of industries rapidly shifted to remote work at the onset of the COVID-19 pandemic. While existing work has examined the impact of this shift on office workers, little work has examined how shifting from in-person to online work affected workers in the informal labor sector. We examine the impact of shifting from in-person to online-only work on a particularly marginalized group of workers: sex workers. Through 34 qualitative interviews with sex workers from seven countries in the Global North, we examine how a shift to online-only sex work impacted: (1) working conditions, (2) risks and protective behaviors, and (3) labor rewards. We find that online work offers benefits to sex workers' financial and physical well-being. However, online-only work introduces new and greater digital and mental health risks as a result of the need to be publicly visible on more platforms and to share more explicit content. From our findings we propose design and platform governance suggestions for digital sex workers and for informal workers more broadly, particularly those who create and sell digital content.
\end{abstract}



\section{Introduction}
The onset of the COVID-19 pandemic led to a rapid shift to remote work across a variety of sectors. While all sectors that were able to shift to remote work experienced significant changes~\cite{brynjolfsson2020covid}, the impact on workers in the informal labor market was especially significant. Informal sector workers, defined as those in informal labor situations such as ``casual or day laborers; domestic workers; industrial outworkers, notably home workers; unregistered or undeclared workers; and temporary or part-time workers''~\cite{webb2020employment}, lack the institutionalized resources available to those in the formal labor sector and thus face significantly higher risks~\cite{cameron2021risky}. Additionally, those working in the informal sector are typically from already-marginalized groups: they are more likely to be women, low income, and/or migrants.

A key question regarding exclusively-digital labor is whether a completely online workplace improves working conditions, particularly for those from marginalized socio-demographic and/or labor groups. Prior work has studied the experiences of office workers in formal labor arrangements switching to remote work during the pandemic~\cite{ford2021tale, bullinger2021computer, pauline2020impact, bezerra2020human, edelmann2021remote, das2021towards, yang2021effects} and the experiences of gig workers (who are part of the informal labor sector) who conduct digitally-mediated in-person work~\cite{cameron2021risky, ravenelle2021side, rani2020platform, apouey2020gig, polkowska2021platform}. However, to our knowledge, no prior work has studied the experiences of informal labor sector workers who switched from in-person work to completely digital work, while continuing to do the same type of work.  By studying this group, we can compare the experiences of those with employer support (formal workers) \add{and (typically) stronger labor rights} with those who lack this support to better understand how to design digital labor tools that inclusively support wellbeing for all.

To this end, we study a particularly marginalized group of workers whose work is typically part of the informal sector~\cite{pitcher2015sex}: sex workers. 
UNAIDS defines sex work as the exchange of money for sexual services or erotic labour~\cite{unaids}. We seek to understand how a shift to online-only sex work impacted previously in-person sex workers' (1) working conditions, (2) risks and protective behaviors, and (3) labor rewards. To answer these questions we conducted a semi-structured interview study of 34 people who shifted from in-person sex work prior to the pandemic to online-only sex work during the pandemic. 
%

Sex workers are highly marginalized due to their gender identity, risk of physical violence, and the stigma associated with their work~\cite{benoit2018prostitution,sanders2018internet}. 
In-person sex work exists under a range of legislative models ranging from criminalisation for seller and buyer through legalisation and decriminalisation
~\cite{smith2018revolting}. 
Online-only sex work (i.e., where the work itself takes place online, rather than just the advertising/scheduling of an in-person meeting) is variously legal, regulated or criminalised under regulations for the making or distribution of explicit or pornographic material~\cite{stardust2018safe}. 
Online-only sex work is a growing labor market sector~\cite{cunningham2018behind} and news reports suggest this growth has accelerated during the pandemic, especially among marginalized groups~\cite{Sexworkp16:online,CammingP33:online}.

The vast majority of our participants shifted to platform-mediated sex work (e.g., selling explicit content via a sex-work-focused online platform) during the first year
of the COVID-19 pandemic. As a result, they drastically changed their digital strategies. Specifically, they: 1) published more information about themselves online, 2) produced and published more, and more explicit, content, and 3) developed a more widespread digital presence (i.e., created more work-related accounts on more platforms and on a wider variety of platforms). In turn they experienced new and/or increased digital threats including stalking and other privacy violations, de-platforming, and content theft. They also experienced greater difficulties with de-humanization and harassment from clients  
as well as with work-life balance, similar to formal sector (e.g., office) workers who switched to remote work during the pandemic~\cite{sandoval2021remote}. 

However, online-only sex work also offered significant rewards. Participants reported a reduced threat of physical violence from both clients and police, including in places where in-person sex work was regulated or legalized. They also reported benefiting from a wider client base which afforded greater freedom of expression (e.g., to look how they wanted without conforming to normative gender or beauty standards) and in some, but not all, cases greater financial earning power. Finally, as in other labor markets, participants reported enjoying the benefits of greater flexibility from online-only work. As a result of these benefits,
the majority (23 out of 34) participants reported planning to continue online work and three reported that they would not return to in-person work at all.

Our findings offer implications both for the policy and platform governance debates surrounding the growing online sex work industry~\cite{blunt2021automating,albert2020fosta,bronstein2021deplatforming,weitzer2020campaign}. \add{Further, we connect these results to }broader conversations on digital labor and its intersection with issues of content moderation, deplatforming, platform transparency, and control over digital content. \add{Finally, we emphasize the role of legislation in addition to digitization in mediating labor rights for informal labor: a critical issue for all gig workers~\cite{stewart2017regulating} but one that is especially complex in the sex industry~\cite{, stardust2018safe, berg2020porn}.}

\section{Prior Work}
Here, we review prior work on digital-mediation of sex work, on sex work (both digitally-mediated and not) during the COVID-19 pandemic, on the shift to remote work across various industries during COVID-19, and on digital safety work in intimate, non-labour contexts.

\subsection{Sex Work and Tech}

The term "sex work" was coined by a sex worker (Carol Leigh) in 1979~\cite{nagle2013whores} and is now widely used to refer to the exchange of erotic services for money. The term includes a variety of activities such as escorting (prostitution), professional BDSM, stripping or lap-dance, phone sex and webcamming. In-person types of sex work are variously criminalised or regulated across the world and in rare cases decriminalised~\cite{smith2018revolting}. Online sex work is generally considered to be making or distributing pornography, which is also highly regulated and in some places criminalised. Sex work is highly stigmatized~\cite{benoit2018prostitution}, and a majority of sex workers are migrants, trans people and overwhelmingly women~\cite{, kempadoo2018introduction, selmi2012dirty}.

Sanders et al. conducted the largest investigation of digitally-mediated sex work in the UK in a 2016 mixed methods study~\cite{sanders2018internet}. Their work focused on how in-person sex work was increasingly being mediated by technology or platforms rather than personal management~\cite{cunningham2018behind}. They also investigated sex worker and client usage of internet fora for various purposes, and law enforcement attitudes towards digitally mediated sex work~\cite{campbell2018technology}. In-person sex workers use the internet to advertise for clients, communicate with them and colleagues, screen out bad clients and in some cases collect deposits or payment. Digital advertising helped sex workers have more control over their work but new kinds of abuse were also tech-facilitated~\cite{sanders2016our}. 
%

More recent \add{prior work has} further explored how in-person sex workers advertise online, strategise for their digital and physical safety, and are discriminated against by digital platforms\add{~\cite{mcdonald2021s,barwulor2021disadvantaged,blunt2020erased,blunt2020posting,blunt2021automating,kuhar2019negotiating}}. 
Additional \add{literature} has explored how sex workers and sex work organizations use technology for social justice and provision of services~\cite{strohmayer2017technologies}, how technology can support sex workers' safety (e.g., through the creation of Bad Client Lists of known violent clients)~\cite{strohmayer2019technologies}, and how sex workers provide virtual peer support and mutual aid to each other via online forums~\cite{barakat2021community, van2018good}. 
Finally, \add{Cowen and Colosi propose that digital sex work platforms be encouraged for safety reasons, through a theoretical "transaction cost framework" analysis~\cite{cowen2020sex}, while} Rand~\cite{rand2019challenging} examines digitally-mediated sex work through a labour perspective and concludes that it constitutes ``sexual entrepreneurship'' and is noticeably absent from the literature on digital platforms. Our work takes a step toward addressing this gap. 

A more limited body of work has focused on online-only sex work. Jones' \textit{Camming} is an in-depth study of webcam workers~\cite{jones2020camming}, and she has also investigated racism in this area~\cite{jones2015black}, and the experiences of fat~\cite{jones2019pleasures} and transmasculine workers~\cite{jones2021cumming}. Stegeman examined the terms and conditions of camming platforms, finding that these platforms attempt to re-frame the webcamming work as ``not-work''
in order to preclude online-only sex workers from worker rights~\cite{stegeman2021regulating}. 
Nayar~\cite{nayar2017working} investigates the way professionalism and amateurism interact in webcamming while Vlase and Preoteasa~\cite{vlase2021flexi} investigate whether camming is a flexible and/or insecure type of platform gig work. Lykousas et al.~\cite{lykousas2020inside} performed a quantitative analysis of a digital sex work platform (Fancentro) just prior to the pandemic, analysing the demographics of performers and their revenue. Abel~\cite{abel2021you} suggests that personal branding and the platforms themselves have a role in upholding class stratification and reinforcing inequality between sex workers. Henry and Farvid analyse online-only sex work from a variety of feminist perspectives and invite further critical analysis of technology-mediated sex work~\cite{henry2017always}.
Swords, Laing and Cook~\cite{swords2021platforms} mapped platform interconnectedness by examining how webcam workers linked  
multiple different platforms to promote their work, attract new clients, get paid and manage their digital identities. In this work, we answer their call to investigate platform use by sex workers during the pandemic. 

\subsection{Sex Work in the Pandemic}
Scholars, particularly in the public health sector, called on governments to include sex work in their coronavirus response~\cite{adebisi2020sex,howard2020covid,jozaghi2020covid,kawala2020policy,lam2020migrant, website:Cis.unipi.it}, including calling for sex workers to be prioritised for vaccination~\cite{vaghela2021sex}. However, many governments did not include sex-worker-specific support in their social service response~\cite{bahadur2020sex, belete2020uncovering, campbell2020global, fedorko2021sex}
. In some countries, funding for organisations working with sex workers was redirected to general coronavirus crisis needs~\cite{lahav2021helping}. As a result, sex workers relied on support from mutual aid (peer monetary support, food banks, PPE drives)~\cite{website:SWARM, website:Sexworkeurope} and sex work organizations (monetary support, harm reduction guides)
~\cite{fedorko2021sex,reza2020community,santos2021sex,dziuban2021very} or pivoting to online-only sex work as was the case for our participants. 
A number of NGOs reported that sex workers were better equipped to deal with the pandemic as compared to the general public, due to their existing health risk awareness and contact with health services~\cite{avwioro2021commercial}, even including the most marginalised such as minors trading sex~\cite{amdeselassie2020experiences,ghimire2021effects}.
\add{In light of the difficulty of collecting primary data in a pandemic, Campbell et al.~\cite{campbell2020global} and Bahadur et al.{~\cite{bahadur2020sex}} conducted media analyses, which documented}
daily life impacts such as movement restrictions, food insecurity, homelessness; violence and exclusion from government schemes; and the response from sex worker rights organisations. 
\add{
Subsequent qualitative research with sex workers echoed these findings: Belete in Ethiopia~\cite{belete2020uncovering}, Janyam et al. in Thailand~\cite{janyam2020protecting}, Museva in Zimbabwe~\cite{museva2021sex}, Pereira in Portugal~\cite{pereira2021male}, Singer et al. in America~\cite{singer2020covid} and finally Tan et al. in Singapore~\cite{tan2021impact}.
}
In some places, the closure or interruption of healthcare services, meaning sex workers struggled to access for example HIV prevention or treatment, was of great concern~\cite{gichuna2020access,macharia2021sexual,toh2020impact, website:Sexworkeurope}, as was the mental health of sex workers dealing with these issues on top of increased stigma and violence~\cite{kahambing2021mental, website:NobodyLeftOutside}. 
Field reports from academics, sex worker rights groups and NGOs~\cite{shareck2021double} reported additional issues such as increased intimate partner violence~\cite{mantell2021life}, forced quarantine for sex workers~\cite{kimani2020effects} or police fines and deportations~\cite{lam2020covid, website:Pion-norge}. Our work builds on these prior findings to focus on the role of online-only sex work in this precarious context.

There is a limited body of prior studies on digital mediation of sex work during the COVID-19 pandemic. Brouwers and Hermann analysed the response of adult advertising platforms to the pandemic, finding various responses ranging from raising money for mutual aid funds for sex workers, to removing valuable safety tools (such as a client review system where sex workers could verify the client was genuine and ask for a reference from a colleague the client had previously seen), in order to appear to not support in-person meeting during stay-at-home orders~\cite{brouwers2020we}. Further, 
\add{Cubides Kovacsics et al. studied} 
the impacts of the COVID-19 pandemic on sex workers \add{in the Netherlands, finding} that the pandemic is ``acting as a magnifying glass for insecurities that sex workers have faced for long''~\cite{cubides2021sex}.

Examining how in-person sex workers were responding to the pandemic, Callendar et al. examined male sex worker advertising profiles during the pandemic and found that fewer profiles were active, fewer new profiles were being made, and some profiles included messaging about the virus including redirection toward online services~\cite{callander2021investigating}. 
Relatedly, Azam et al. described changes in the in-person sex work market, comparing the Netherlands and Belgium, neighbouring countries with differing laws regarding sex work, finding a shift from venue (brothel, parlour or club) work towards independent sex work (escorting)~\cite{azam2021covid}. However, this study uses client reviews and sex worker advertising directories and forums as factual data, an approach that is controversial due to the likely low veracity of this data~\cite{website:forgedintimacies.com}. Our work addresses their call to investigate ``what happens to sex workers who stop selling physical sex,'' and uses qualitative methods to deeply understand workers' direct experiences of the impact of a shift to online-only work. 

\subsection{Remote Work in the Pandemic}
In the US, as many as 1/3 of workers switched to remote work, and a further 1/10 lost their jobs or were furloughed in the first months of the pandemic~\cite{brynjolfsson2020covid}. In Europe, 
\add{Ahrendt et al. estimate}
the prevalence of remote work as high as 50\%~\cite{ahrendt2020living}
\add{also reporting that}
those switching to remote work were more likely to be younger people and women\add{: a finding that was echoed by Brynjofsson et al.~\cite{brynjolfsson2020covid}.}

A sizable and growing body of literature has investigated how the shift to remote work affected office and professional workers' productivity~\cite{bezerra2020human,pauline2020impact,bullinger2021computer}, collaboration approaches~\cite{yang2021effects}, the accessibility of their workplaces~\cite{das2021towards}, their work-life balance~\cite{sandoval2021remote}, stress~\cite{sandoval2021remote}, digital fatigue~\cite{bennett2021videoconference}, and degree of worker surveillance~\cite{aloisiessential}. Broadly, this prior work finds reduced collaborative efficacy, reduced work-life balance, increased stress and digital fatigue, and increased workplace surveillance. Despite these consequences, remote work offers benefits such as improved productivity and job satisfaction, greater accessibility and flexibility, and prior work finds that a remote workplace appeals to workers and was a key criteria in their pandemic job searches~\cite{edelmann2021remote,mcfarland2020impact}. 

The benefits and consequences of the pandemic switch to remote work have not fallen equally. Women~\cite{galanti2021work,brynjolfsson2020covid, sandoval2021remote,ahrendt2020living}, migrants~\cite{giorgi2020covid}, and those whose jobs prior to the pandemic involved contact with the public~\cite{giorgi2020covid} were the most negatively impacted by the shift to remote work. These negative impacts included reductions in mental health and income, and for women in particular, higher burdens of childcare and housework. 
Our work responds to these findings by investigating 
whether and how the changes the pandemic forced affected a particularly marginalized population made up of exactly the young, migrant, mostly women in contact with the public that these studies have shown were most affected by COVID-19 in formal workplaces. Beyond workers in the formal sector, prior work also examines the experiences of gig workers who may conduct digitally-mediated but still in-person work during the pandemic. Gig workers such as those advertising to do home-based tasks on Taskrabbit faced significantly more risk and challenging tradeoffs than professional workers: they were forced to chose between the risk they were prepared to take with regard to COVID-19 and financial wellbeing~\cite{cameron2021risky}. To navigate this tradeoff, as well as clients expectations of their behavior, they developed a number of strategies (altering their pricing for high risk tasks, engaging in safety theater such as wearing more protective gear than they otherwise felt they needed to in order to reassure clients) and in some cases stopped gig work entirely. 
However, the ability to stop working was a privilege. Like many sex workers, other gig workers experienced many barriers to receiving government aid and therefore had no choice but to continue working through the pandemic - in this way we can view gig work (including sex work) as a non-governmental financial safety net~\cite{ravenelle2021side, apouey2020gig}. Digitally mediated in-person workers were often also ineligible for healthcare, and could not social distance (e.g., ride-share drivers)~\cite{rani2020platform}. 
\add{Ravanelle et al. and Stephany et al. both}
found that, as with sex workers, there was a growth in workers entering the digitally-mediated gig workplace due to the pandemic and because of the ``distancing'' effect - i.e., more goods and services being distributed via digital mediation - that there was the possibility of increased income during this time~\cite{polkowska2021platform, stephany2020distancing}.

The vast majority of prior work on remote work during the COVID-19 pandemic focuses on employees shifting from in-person to remote work \textit{or} on gig workers continuing (or beginning) digitally-mediated in-person work during a pandemic. In contrast, our work focuses on a primarily-informally (gig) employed workforce shifting to completely digital employment. Further, we focus on a workforce composed of the groups prior work finds have been most affected by the shift to remote work that is additionally stigmatized and marginalized due to the nature of their work. By focusing on a group of workers at the intersection of multiple dimensions of marginalization and precarity we can offer insight into the promise and peril of remote work for this group and other groups of workers who intersect with them on one or more axes.
\add{\subsection{Digital Intimacies}}

\add{Prior work also explores safety work in intimate digital contexts outside of labour~\cite{gillett2021not}. Prior work investigating people's experiences with online dating find that app daters (like sex workers) seek control over their personal information, including whether and how they linked to their private social media and how much information was shared about them with potential matches~\cite{albury2019safety,cobb2017public}. For safety purposes,  online daters also use some of the same protective strategies as in-person sex workers, such as using personas with fake birthdays or names in the early stages of dating~\cite{obada2017don}, vetting potential matches before meeting them~\cite{gill2012violence, cobb2017public,pruchniewska_i_2020,obada2017don}, ``covering'' -- a practice of sharing your location with a trusted contact or friend while you are with the stranger from the app~\cite{byron2021hooking} -- or blocking those who send them harassing messages~\cite{albury2019safety,pruchniewska_i_2020}. Further, Stoicescu and Rughinis considered the actual data practices of online dating websites, concluding that users had a greater right to security assurances than currently exist, in view of the fact that appearing to engage, even in non-commercial intimacy, was a reputation risk~\cite{stoicescu2021perils}.}



\add{Additional prior work has explored the digital threats faced by those in relationships, in situations of intimate partner violence (IPV) finding extensive use of ``spyware'' technologies as well as digital stalking by abusers~\cite{chatterjee2018spyware,dimond2011domestic,matthews2017stories}. IPV survivors who have left an abusive situation and seek to avoid discovery by their abusers and sex workers use many of the same privacy protective strategies including using multiple devices and/or personas, as well as limiting what personal information or media they share online~\cite{matthews2017stories,dimond2011domestic,andalibi2016understanding}. Interventions to address these harms include "Clinical Computer Security" services for the victims~\cite{havron2019clinical, freed2019my}. 
Such an approach would be life-changing for sex workers experiencing digital abuse, and could even be sponsored or implemented by platforms themselves.}\\
%
%
%
%
%

\section{Methodology}
\begin{figure}
    \centering
    \small
    \includegraphics[width=0.7\textwidth]{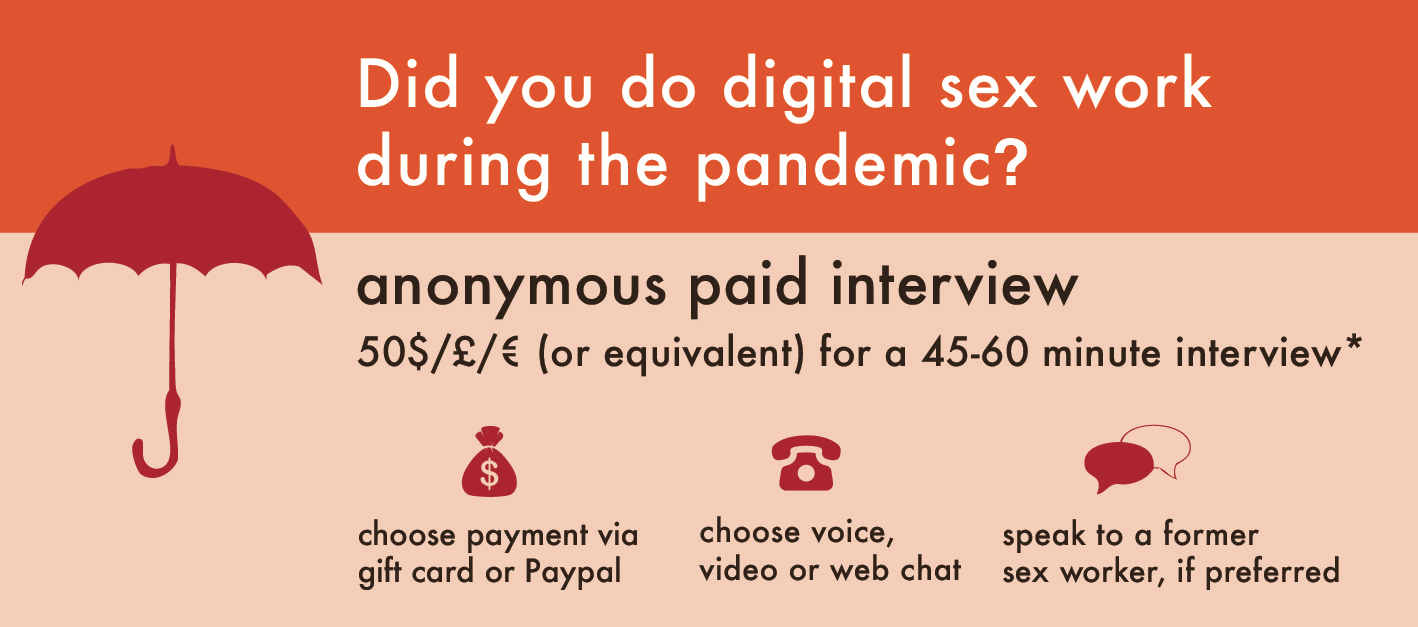}
    \caption{Recruitment flyer.}
    \label{fig:flyer}
\end{figure}

We conducted semi-structured interviews with 34 participants who had done in-person sex work prior to the pandemic. The kinds of work they reported doing included stripping, escorting, studio porn and professional BDSM. Because sex workers are a hard-to-reach population~\cite{shaver2005sex}, we did not limit our scope to any particular country, however recruitment was so successful that in hindsight limiting to one country would have been possible. Recruitment was done by sending a recruitment graphic (see Figure~\ref{fig:flyer}) and sign-up link to well-connected sex worker contacts of the researchers and asking them not necessarily to participate in the study themselves, as this would limit the results to only sex workers connected to the researchers, but to share it widely~\cite{biernacki1981snowball}. The sign-up link led to a Qualtrics screening survey, which screened out minors and people who had not done in-person sex work prior to the pandemic. Qualified participants read and signed the consent to participate as part of the screening process. The screening survey ended with a link to Calendly, an online service where participants could anonymously self-schedule an appointment for an interview. Interviewees were paid \$50 dollars or local equivalent and could choose to be paid by PayPal or with a voucher (e.g., Amazon gift card). 

Interviews ranged in length from 27 minutes 18 seconds to 85 minutes 33 seconds, with an average of 53.6 minutes. We used an end-to-end encrypted Webex interview system, which ensures that participants do not need to sign in or give any identifying information to participate. Interviewees gave consent a second time, verbally, to be interviewed prior to starting the interview. Demographic information was collected by the interviewer after the recorded part of the interview had finished and noted separately (see Section~\ref{sec:meth:descriptives} for sample demographics).

We began the interview by asking about the participant's sex work background and the kind of work they were doing prior to the pandemic. We then asked them to describe any kinds of online-only sex work they were doing. The protocol probed their experiences with online-only sex work, as well as the risks they were facing. There were specific questions about finances as well as the participant's view of their future career choices. We also asked questions about any benefits they might have experienced from doing online-only sex work and what their clients were like. 
Finally, we queried participants about which platforms they used and how they interacted with the platforms, as well as more generally what tools and strategies they used and what did not work for them. The interview questions analyzed in this paper are presented in Appendix~\ref{appx:protocol}. 

Interviews were transcribed and then qualitatively coded using MaxQDA coding software. \add{We used an inductive, open-coding process that most closely resembles the Grounded Theory method~\cite{muller2010grounded}. As a first stage of analysis, themes that presented in four randomly selected transcripts were organised into a codebook.}
Two researchers then coded eight transcripts, iteratively updating the codebook as needed, checking for and achieving agreement of 82\%.  
Given this high degree of agreement, the remaining transcripts were then coded by the first author. 
\add{After coding all the transcripts, the first and third author selected semantically proximal themes to report: this paper is the first analysis of these themes.
Quotes that succinctly, descriptively and meaningfully illustrate each finding were drawn from the transcripts using the coded segment function in MaxQDA.}
We primarily report our results qualitatively, using terms such as many, few, and some~\cite{mcdonald2019reliability}. As an exception, when providing overall context (e.g., on the types of work participants performed) we offer counts to anchor the reader's perception. We emphasize, however, that our work is qualitative in nature and should be interpreted as such.

\begin{figure}[t!]
     \centering
     \begin{subfigure}[!t]{0.43\linewidth}
         \centering
        \includegraphics[width=\linewidth]{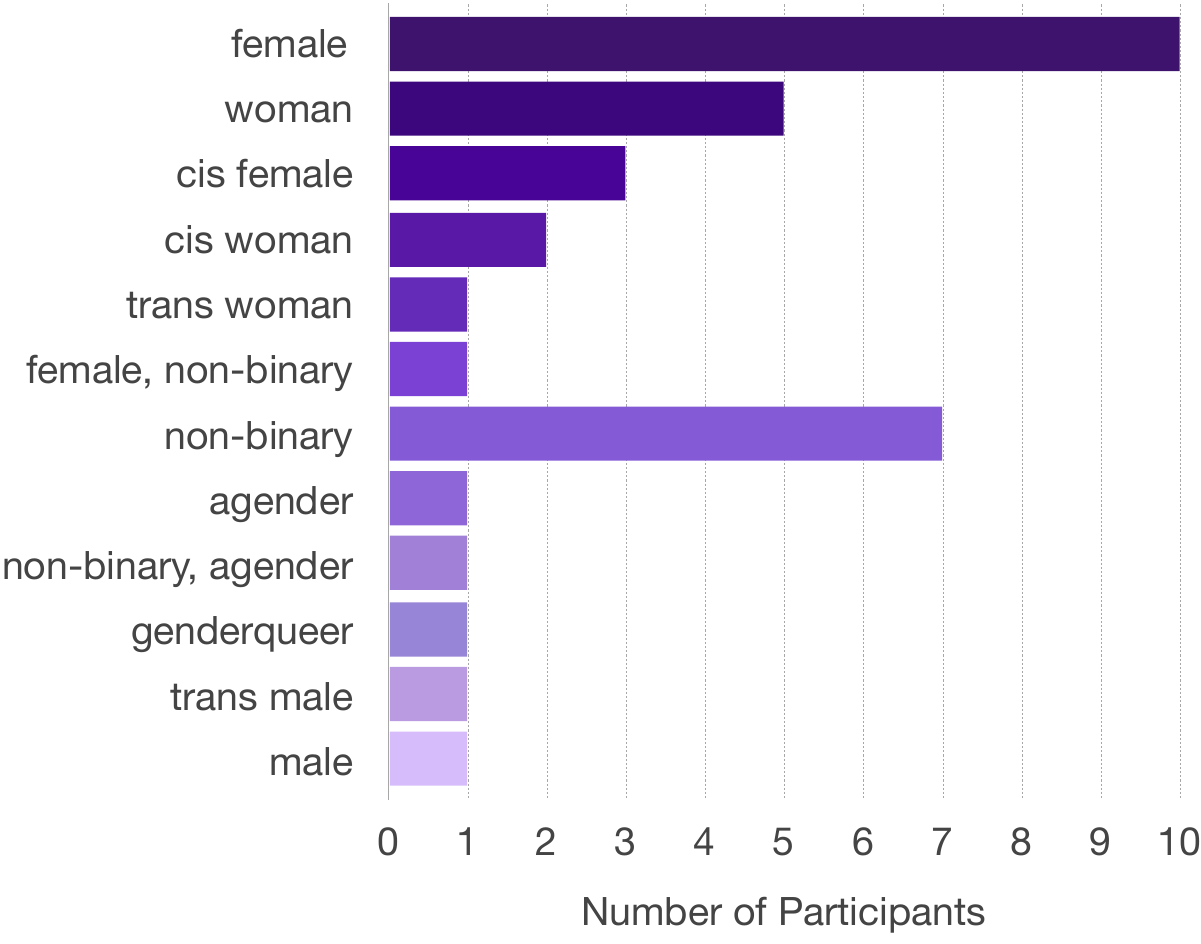}
        \caption{}
     \end{subfigure}
     \hfill
          \begin{subfigure}[!t]{0.56\linewidth}
         \centering
       \includegraphics[width=\linewidth]{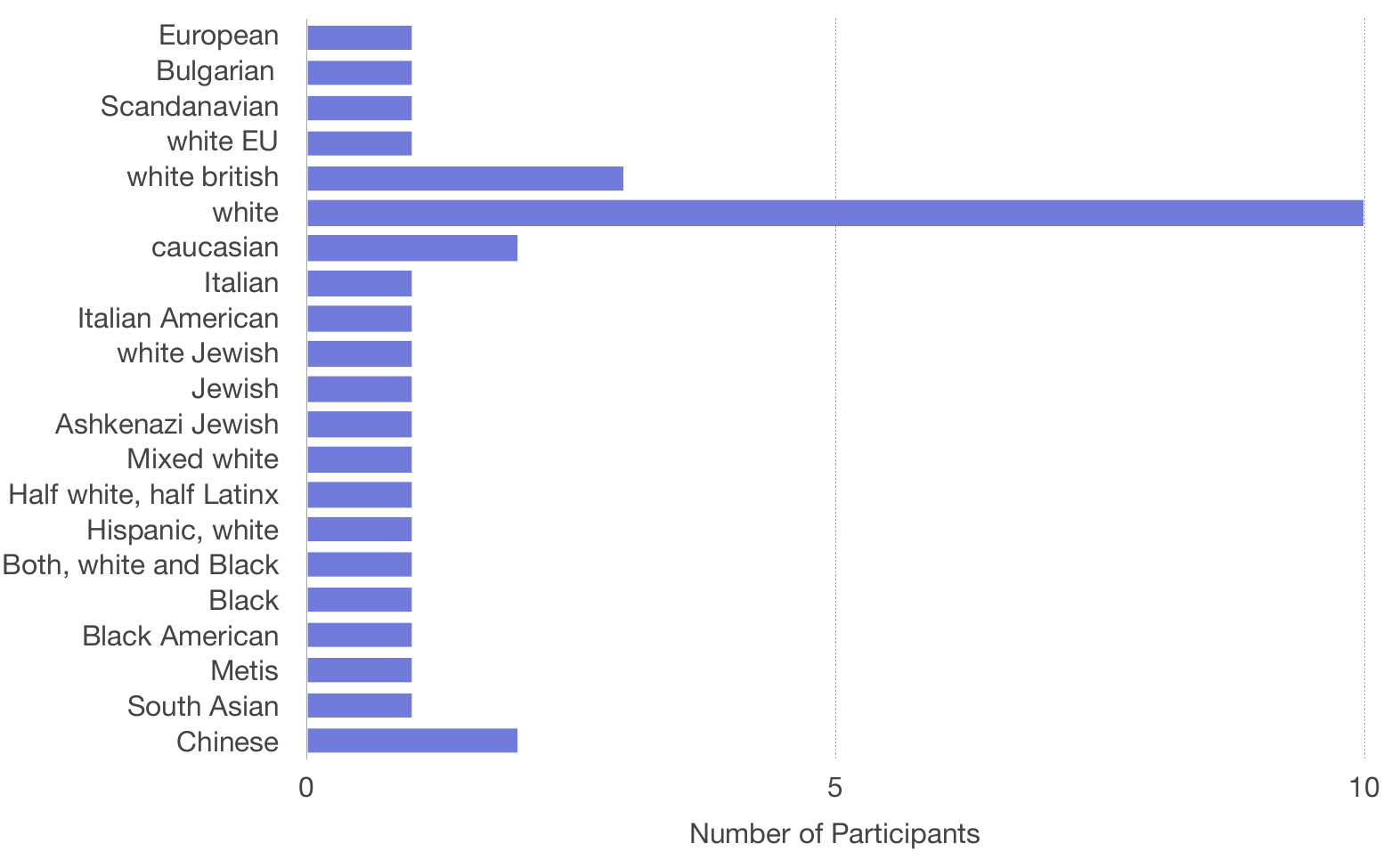}
        \caption{}
     \end{subfigure}
     \caption{\add{Self-reported participant demographics. On the left, gender (a) and on the right, ethnicity (b).} 
     }
         \label{fig:demographics}
     \end{figure}

\subsection{Sample Demographics}
\label{sec:meth:descriptives}
\add{All of our participants primarily offered direct-to-client in-person services prior to the pandemic.} Participants had a mean age of 29.3 years, with a standard deviation of 4.23 years. Participants worked in seven countries: the USA (11), the UK (11), Germany (5), Sweden (3), Spain (3), France (2), and Canada (1); some participants worked in more than one country. Half of our participants (17/34) were disabled, however few of them disclosed this in the context of their sex work. All 34 participants described themselves as LGBTQIA, however 10 participants reported that they either hid their sexuality at work, or used a different one to their private identity. Participants were invited to self-describe their ethnicity and gender. As such, we report this data as they described it \add{in Figure~\ref{fig:demographics}}.

\add{Sex workers are more likely than the general population to be LGBTQ~\cite{selmi2012dirty, laing2015queer, smith2012introduction, shah2012sex, website:nswp.org} and/or disabled~\cite{fritsch2016disability}.}\footnote{\add{Measuring the size and demographics of the sex working population is a very difficult problem~\cite{cusick2009wild}. For some estimates of the demographics of the sex working population, see \url{https://tgeu.org/faq/dec-17/} and \url{https://survivorsagainstsesta.org/lgbtq/}.}} \add{Whilst the reasons for this are complex and not well-understood, they are likely to include a lack of familial support in youth, homophobia and ableism in education and the workplace and other factors. Additionally, LGBTQ and/disabled sex workers may be more precarious and therefore more likely to participate in paid research of this kind.} 
%

Finally, we asked participants to share what one hour of their time cost to the client for an in-person session prior to the pandemic. Some participants offered sessions in shorter intervals, in which case they pro-rated their costs. These amounts ranged from £50.16 to £1471.83, with a 
\add{median cost of £200}
and standard deviation of £257.40. These figures must be viewed with the understanding that participants offered varying services, in different countries, under varying legislative regimes. \add{The high standard deviation can be explained by outliers: one participant offered readily available services in an environment where there is a high supply and low labor wages across most sectors, while another participant offered a niche service in a criminalised environment in a wealthy city.} 

\subsection{Ethics} 
Our research was approved by our institution's ethics review board. We took care to take extensive measures to preserve participant anonymity and data protection at every stage of the research process. Webex and Calendly were used specifically to avoid participants' having to give any identifying information. Guidance to set up a ProtonMail account was shown in case participants opted for a payment method (PayPal) that required an email address to receive payment. 
At the start of the interview, the interviewers explained carefully the general aims of the study (taking care not to prime the responses) and gave participants an opportunity to ask any questions before the recording was started. Some participants asked about data privacy and anonymity and were told that the study adhered to both the Institutional regulations and GDPR and that the researchers' declared intention to take the utmost care of the identities and information received, in view of the participants' marginalized, stigmatized, and potentially criminalized work. Participants were also invited to not answer any question that they did not want to answer. 

In line with prior work~\cite{barwulor2021disadvantaged,barakat2021community,razi2020let}, we do not refer by name to platforms and tools that our participants used for sex work\footnote{There are two exceptions throughout the paper. As a general example, we mention the platform OnlyFans, as it has received significant media attention and is widely known as a platform that mediates online-only sex work~\cite{SexWorke70:online}. We also mention an example of a class of tools in the Discussion but we specifically chose a tool \textit{not} explicitly mentioned by our participants.} in order to protect the sex workers using them. We have also anonymized place names and any other identifying information.

Finally, in line with research justice best practices ~\cite{macneil2006informing}, we will send participants who requested a copy of the final paper. Additionally, we plan public science communication efforts around this work, which will feed back results of the study to the communities involved. \add{These will include mixed-media presentations of research results in accessible, actionable language (e.g., webinars, infographics, one-page results summaries). These resources will be disseminated through the authors’ existing relationships with sex worker rights groups as well as promoted through the channels used to recruit participants. 
Lastly,} 
as an added component of research justice, we chose to employ a sex worker with transcription skills to perform the transcription for this research. 

\subsection{Positionality} The authors are scholars of sex work and technology. Sex workers were centred at every stage of the research. However due to the small number of researchers and the large number of identities represented in the research participants, not every participant would see themselves reflected in the research team. We particularly note the lack of Black, Asian and Latinx researchers on the team, as we are white/Brown and Arab-American/American/british\footnote{Authors prefer not to capitalise british as a decolonization practice~\cite{website:mtroyal}.}. Additionally there were male study participants and no male researchers, which limits our ability to interpret that data to the fullest extent possible.

\subsection{Limitations} 
Every stage of this study involves limitations. The interviews were only conducted in English, which limited the sample to English-speaking participants. Chain sampling~\cite{biernacki1981snowball}  
avoided the possibility of people who were not suitable signing up for the study, however it limited recruitment to people who were in the networks of the sex workers connected with the researchers, and their wider circles. However, the recruitment graphic organically reached several large private social media groups where sex workers communicate with peers; this widened participation considerably. A concern with this type of recruitment is that the participants might all have similar experiences. Accordingly, we note that the study was Western-centric and that we received participation only from European and North American residents, which limits the generalizability to other regions. However, among our Western participants, both the sample demographics and the wide variance in perspectives reported suggest that we captured a wide spectrum in the experiences of those engaged in sex work in these regions.

\section{Results}
In this section we summarize our results. We first provide as context a description of the types of work our participants performed. We then describe their working conditions and work experiences, the risks they encountered and how they protected against those risks, the rewards they found from this work, and whether they planned to continue online-only work in the future.

\subsection{Types of Work}
Almost two thirds of our participants (20/34) had done some kind of online-only work before the pandemic, although this was not necessarily the type that they chose to switch to at the onset of the pandemic. The most common situation\footnote{Ten of 34 participants had previously done webcamming; the second most common was phone sex, which 4 of 34 participants had done.} was to have tried or undertaken a period of platform-mediated web-camming -- in which workers perform virtual shows ranging from strip shows through explicit porn-style performances~\cite{jones2020camming} -- at some point within their sex work career as a whole. \add{For example, }P22 described that they:
\begin{quote}
``tried webcamming but it's too demanding for me and you can sit there for hours and not make anything and that's not worth it for me. I did it when I started [sex work], but I'm tired now.''\end{quote}

After the onset of the COVID-19 pandemic, our participants broadly undertook three types of online-only sex work: a) exclusive arrangements;  b) direct-to-client sales; and/or c) platform-mediated work. 

\subsubsection{\add{Exclusive Arrangements}} 
Exclusive arrangements, where a single client would engage the worker privately for constant or very regular contact, were the rarest type of online work, probably because they involve a very significant amount of money that most clients were unwilling to spend. 
\add{Additionally, such arrangements involve a significant amount of both affective and emotional labor~\cite{makinen2021resilience}. As P27 described, because they now depend on a single client, they have had to make ``a lot of compromises that [they] wouldn't have before.'' These compromises were ``especially in terms of time, I feel like I'm always on call now, like a boyfriend or something.''} 
Only two participants were in exclusive arrangements. One other participant described having a small number of high-touch arrangements with high value clients as a way to maintain contact with existing clients but avoid in-person COVID risk, as also described by prior work~\cite{azam2021covid}. 

\begin{figure}
    \centering
    \small
    \includegraphics[width=0.8\textwidth]{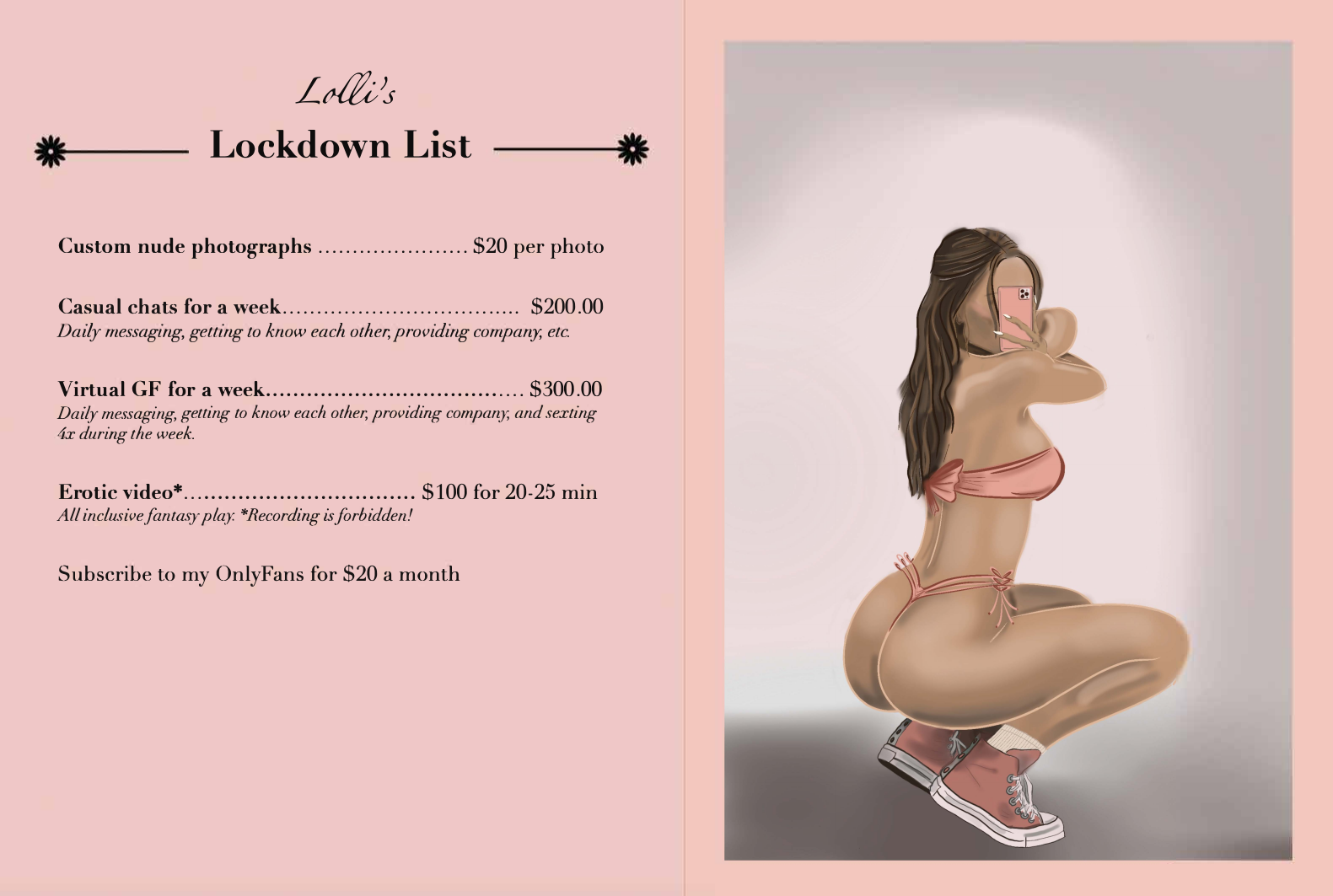}
    \caption{Stylized example of a service and content menu that a worker might use to advertise Direct-to-Client sales on social media.}
    \label{fig:dtc}
\end{figure}

\subsubsection{\add{Direct-to-Client Sales}} 
Twenty-six participants did direct-to-client sales. Direct-to-client sales involved offering some or all of a very wide range of products including pictures, videos, phone calls, texts, texting packages, sexts (sexual texts), emails, \add{audio recordings, body-part rating videos}, video calls and private gallery access.


Direct-to-client (DTC) sales have the benefits of receiving the full profits from their sales (platforms that mediate content \& services sales for sex workers, see further discussion below, generally take a percentage from sales of around 20\%) and not requiring any additional account set-up, which might involve submission of banking details or government ID. However, this style of work involves a great deal of administrative work, a stable and adequate (i.e., anonymous for the worker) payment processing account, and content is not protected from being downloaded/captured and shared externally (we discuss the latter threat further in Section~\ref{sec:results:risk} below). Participants engaged in DTC sales generally advertised this using the same strategies that they had previously advertised their in-person work, which included posting on public fora and social media, posting on directory listings, maintaining their own email lists through which they sent newsletters, and posting on their own websites. Many participants started making small social media graphics that they posted on their existing (or newly created) social media accounts with a menu of their DTC offerings (see Figure~\ref{fig:dtc} for a stylized example).

During the first few months of the pandemic, some participants did not yet anticipate the length of time in which they might be unable to work in-person, and offered DTC sales rather than taking the further step of signing up to a sex-work-specific content creation platform. However, 7 of our participants switched from DTC sales to platform work during the pandemic. P26 explains that:
\begin{quote}
``At the beginning of the pandemic I was doing video calls off any platform with clients I already knew, but now all of my work is [via subscription content site].'' (P26)
\end{quote}

\subsubsection{\add{Platform Mediated Work}} 
The vast majority of participants (27/34) did platform-mediated work; 8 did only this work, while 20 did both platform-mediated work and direct-to-client sales.

An enormous range of platforms were used for platform work, including membership/subscription sites: ``I'm currently doing [subscription content site], so that's explicit videos and photos for a monthly subscription cost.'' (P2); clip sites, where creator-generated content is hosted and sold; camming (see earlier definition), phone sex (explicit voice calls~\cite{selmi2012dirty}) and sexting (the exchange of suggestive text messages~\cite{weisskirch2011sexting}) platforms, where the client can contact a sex worker for these services; and sites whose primary focus is in the sale of custom (client-requested) content and physical personal items such as used underwear~\cite{sanders2018internet, swords2021platforms}. 
 
In addition, some multi-modal platforms where workers had advertised in-person services with some additional content for sale added new affordances to enable online-only sex work as part of workers' existing in-person advertising profiles. For example, platforms added functionality to process payment for online services, conduct video calls, or advertise digital services.

In addition to using online sex work platforms, participants used social media much as other entrepreneurs: to build their brands, advertise for new clients (and direct them to the online sex work platforms through which they offered content and/or services), as well as to network with peers~\cite{swords2021platforms}.  
%
Participants chose, or switched to platform-mediated work because of the platform affordances: internal payment processing, passive income opportunities (where the sales are made without direct intervention from the sex worker), market share, convenience, age-verification of clients and content control. We discuss the benefits of these affordances and where existing implementations fall short in the following sections.

While many of our participants chose a particular type of online-only sex work based on personal and business preferences, a few 
found their choices restricted due to lack of access to adequate technology (hardware, wifi, etc.). For example, P9 explained \add{that ``in the beginning I had a different laptop that didn't work that well, so I couldn't do camming or anything like that because I couldn't download anything.''}

\subsection {Working Conditions \& Experience}

Our participants reported that online-only sex work differed from in-person sex work not only in the way that it was delivered, distributed and paid for, but also in the subjective experience of the work.

\subsubsection{\add{Client Behaviour}} 
Participants reported that online-only clients acted more entitled and more likely to engage in harassment, whether verbal harassment or pushing boundaries (i.e., asking for services or content 
that the provider had already clearly described that they did not offer). \add{For example, P1 describes their clients as ``much more cheekier and persistent'' and found that it was ``harder to talk to clients for online work, because negotiation was a lot longer and needed a lot more patience.'' Some participants attributed this difference in behavior to the anonymity of online interactions.}
%

Participants reported starting to be clearer and more direct in their boundary-setting, ignoring requests which fell outside their comfort zone and using the safety structures built into platforms (i.e., blocking) as a last resort. In addition to blocking clients for harassment\add{, as reported in prior work on both work and recreational digital intimacy~\cite{albury2019safety,pruchniewska_i_2020,mcdonald2021s,sanders2018internet},} workers also blocked clients who behaved in ways not allowed on the platform. For example, even alluding to an in-person meeting is banned on many subscription content sites for legal reasons, so blocking clients who mention it is a safety strategy to avoid platform loss, as we discuss further in Section~\ref{sec:results:risk} below.

\subsubsection{\add{Burnout}} 
Mental health struggles were experienced by many workers across different sectors during the pandemic for reasons such as lack of space to work at home, no access to friends or family, upset to routine, lack of privacy, income instability, etc~\cite{hayes2021perceived, jimenez2021impact, galanti2021work, wang2021achieving}. These concerns were also present for the participants in our study. As one participant put it when asked whether they felt burnt out: ``All the fucking time'' (P32); many participants agreed. 

Participants struggled with work-life balance for four primary reasons. \add{ First, workers felt like they needed to be online constantly to respond to clients. Second, and relatedly, workers felt they could not take a break or vacation for fear of not earning income. In order to get these breaks they felt they needed to ``make up excuses or say I was out of signal to try to [get] some breaks and time'' (P27).} 

\add{Third, workers felt pressure to constantly look for opportunities to create content (e.g., while engaged in daily life activities). For example, P1 describes that they ``ended up always being on the lookout for opportunities for filming, even if you're out doing something else you might think, oh this would look good, this might sell, I feel like my mind was always on it and I ended up focusing so much on it.''}

\add{Fourth, workers experienced altered perceptions of themselves due to time spent online (e.g., viewing other people's edited or stylized bodies). For example, P22 explained that they ``struggle more with body dysmorphia now'' and P26 explained that:}
\begin{quote} ``it feels really bad to be on the internet all the time. I can't explain it, there's been moments when I'm on [social media site] and [subscription content site] all day and I will feel so insecure and bad of myself, I feel like it's [because I'm] constantly looking at images of women...even though...I'm not feeling like I'm comparing myself to people in a conscious way. I've sometimes gone out for a walk after being online and looking at real people has been so reassuring...there [is] something about constantly looking at a screen and being on a digital platform and looking at this constant influx of images that have been manipulated, posing a certain aesthetic or dynamic that is quite damaging and exhausting. It makes you feel alienated and drained from real life.'' (P26)
\end{quote}

Our participants responded to burnout by decreasing their workload, using scheduling tools either internal or external to the platform to release content when they themselves were not live on the platform, hiring assistants and simply stepping away from technology and taking time out. 

\add{P13 explained that their boyfriend helped them in their work, ``he was the one sexting them, he was the one scheduling everything, so I wouldn't have made 1 dollar if it wasn't for him.''  In addition to support from their boyfriend, P13 eventually hired an assistant who suggested that they get coaching for handling burnout; these efforts helped them feel ``more centered'' but they explained that ``burnout is like really easy to have with online work.''}
%
%


Whilst burnout is detailed in prior work on digital entrepreneurship in sex work~\cite{rand2019challenging} and in sex work more widely~\cite{del2021cisgender}, our participants described that the additional mental health burden and financial precarity of the pandemic added to their stress. \add{P5 explained that their typical coping strategies, e.g., ``getting a massage or going for a hike,'' were difficult to implement during COVID. Additionally, their burnout was exacerbated by the fact that the pivot to online work meant that they had to work more than before in order to earn enough income because they ``just weren't making as much, it's like [only] a [little] bit of money every day so I had to do it more consistently.''}

\subsection{Risk}
\label{sec:results:risk}
Here, we review the primary categories of risks encountered by our participants in their online work and how they aimed to mitigate these risks, when possible.

\subsubsection{\add{Platform Loss}} 
Platform loss -- usually due to account deletion -- is the single most important risk that participants reported facing in the course of their work: all of our participants discussed this risk. This could be loss of online sex work platforms through which workers offered content and services, loss of social media or other accounts through which workers advertised content and services, or loss of accounts that workers used for DTC sales.

Since some of the workers relied entirely on one platform, platform loss would result in complete loss of income. One participant described how the risk of platform loss affected their mental health:
\begin{quote}`I'm trying to find other jobs, I don't want to depend on [subscription content site]. \add{[The platform]} is basically this thin ice I'm walking on, and at one point it's going to break, and everything's going to go to [expletive], and I'm going to fall through.'' (P30) \end{quote}

Platform loss frequently occurs due to failure to comply with terms of service, which change frequently in relation to explicit content on both sex work and non-sex work platforms~\cite{mcdonald2021s,blunt2020posting}. To mitigate these risks, sex worker scholars such as Sophie Ladder~\cite{website:sophieladder} and Amberly Rothfield~\cite{website:amberlyrothfield.com} intensively analyse the differences in terms of service between platforms and communicate what sex workers are and are not allowed to include in their content, bios and messaging. Participants described being highly attentive to the Terms of Service despite the fact that content moderation, as previously documented, was not equitably or consistently applied, and frequently regulates identity as a sex worker, not just sexual content~\cite{blunt2020posting,barwulor2021disadvantaged}. Complicated navigation of each platform\add{'s rules} was a common theme in the interviews. \add{Participants described how they ``don't use hashtags because they can get you shadowbanned\footnote{Shadowbanning is a content moderation technique where a user can post but the content is hidden from or rarely shown to others~\cite{blunt2020posting, are2021shadowban}. The user is not notified and may not realise they have been shadowbanned until they experience loss of engagement or income.}'' (P4), try to carefully track the ever-changing rules about what they can and cannot post on different mainstream and sex work platforms, and even changed their account settings or persona (e.g., gender, age) ``in the hopes [their account] won't be seen as inherently sexual'' (P25) or to avoid biased content moderation.}

Generally, participants resorted to self censoring and developing alternate language to still communicate with their clients in messaging, material and captions in an attempt to avoid platform loss. \add{For example, some platforms don't allow users to mention external websites like Skype or Paypal, so P30 explained that workers ``just find a way around it by typing other letters.'' Some participants also described using tools that would predict whether their content would be censored so they could avoid posting content that might result in them losing their account. P13 explains one such tool:}
%
\begin{quote}``There's a page where you put a photo and [using] the same [ML] that [social media platform] uses... [it] tells you [if] it's too sexy... if I can upload this to [social media platform] or not.''
\end{quote}

Some participants reported nesting their accounts -- linking increasingly explicit accounts together -- to ensure they were still visible on public social media and could direct their clients to their explicit content while avoiding censorship. \add{For example, P21 explained their four-layer nesting process: that from their social media page they``direct people to my [other social media] and from [other social media] I direct them to my page that has no 18+ links and from there [to] another page that directs them to the 18+ links.'' }
%
However, such nesting is not ideal as some potential clients drop off at each step of the sales funnel.
%

Despite participants' best efforts, their accounts were still frequently deleted on both sex work and non-sex work platforms. Participants constantly threat modelled their risk of platform loss and employed a number of different strategies to plan for it. One tactic was client relationship management: making lists of client email addresses, warning clients that platforms change and deletions happen and taking ownership over client information preemptively.
\add{P24, worried about account deletions, made such a list, telling clients "make sure I've got your email address so I can find you if I need to"}


Participants also used strategies such as VPNs and new devices when accounts were deleted, as many platforms will not allow a new account to be created from the same digital location. \add{For example, P17 described how they ``just made 2 new [free content sharing site] accounts on [2] phones with VPNs always on'' when two of their previous accounts were permanently banned.}

Some participants preemptively made accounts on platforms they did not actively use, planning for the inevitability of needing to switch from one platform to another. 
\add{For example, P31 described taking the username for their own use in this way "whether I use it or not I make an account so nobody else can use my screenname...just in case I want to use it at some point."} 

In addition to the risk that workers will lose their account on a particular platform, there is always the risk that a sex work platform will be shut down entirely as VC funding and financial services necessary for the operation of sex work platforms is often inaccessible or precarious~\cite{SexWorke70:online}. The lack of stability in sex workers' independent access to payment processing and banking is well-documented~\cite{mcdonald2021s, blunt2020erased}. Participants viewed the platforms and apps where they were engaged in online sex work as similarly fragile, and they withdraw immediately or regularly any income generated within the platform to avoid income loss in case of account deletion.
\add{P23 explained their concerns about having "money stolen by these platforms" since the platforms often "shut your account, leave your money in limbo and typically take it." P30 added that they withdrew their money from platforms "whenever it's there, whenever it's ready.'' As a result, they ``make withdrawals all the time'' to ensure they don't let their money ``sit there,'' in the platform's control.} 
%
%
Indeed, in August 2021, one platform that mediates digital sex work, OnlyFans, announced -- and later reversed -- a decision to ban all explicit content sales, reportedly as a result of funding pressure~\cite{SexWorke70:online}. 

\subsubsection{\add{Greater digital exposure}}
In order to effectively market their services and offer content clients were willing to purchase
, our participants found that online-only sex work required them to be ``more online'' and more explicit online.

Participants explained that online-only sex work meant revealing far more intimate details about their work and themselves, which could in turn be used to hurt them. 
\add{P15 captured this feeling succinctly, explaining that ``even if we're being careful there's always a risk to having your body on the internet that much.''}
%
%

Specifically, being ``more online'' increased participants' risk of doxxing, outing (having others in their life discover that they did sex work), and stalking. 
%
Stalking is a concern for all sex workers, as documented in prior work~\cite{gill2012violence,mcdonald2021s}. However, for those engaged in online-only sex work, the breadth of their audience also expanded their potential stalkers. One participant described a particularly chilling incident: 
\begin{quote}``I took a selfie in the bathroom and I upload[ed] that one to [social media site], there was a guy that went to the same bathroom and posted that selfie of him in that bathroom under mine, like I know where you are.
'' (P13)
\end{quote}

Participants used a range of digital strategies to cope with these risks. They took great care to remove visual context from their content that could reveal their identity or location. For example, they avoided creating content in spaces with natural light so it is not possible to tell the time of day, threat modeled whether the shape of their room could reveal their location, and/or posted photos out of order or once they had already left a particular location. 
\add{For example, P11 described switching off location tagging for photos on their phone, noting that they "never post a photo of [themselves] that's outside [their] house or any distinct location." P12 describes using old photos to claim they were somewhere other than their true location, while P31 similarly described posting pictures out of chronological to avoid stalkers, saying "since I take a lot of pictures in hotels, I make sure I never post them until I'm fully sometimes out of the state...it's too easy to take a selfie and post it online but it's better for me to sit on it and make sure I protect myself."}

Participants were also careful to keep their work and personal lives separate to avoid potentially dangerous context collapse, such as by using physical strategies like separate devices (as noted in prior work~\cite{mcdonald2021s}). 
\add{P33 explained, "I have a phone for my personal contacts and a phone for my work contacts, I have different emails", but also expressed privacy despondency over their lack of ability to fully separate their personas: "in the end a lot of the information gets tangled because I log in on the same laptop."}

They also used digital strategies such as having different accounts, as P33 describes above, and utilizing platform affordances that allow workers to block clients from the worker's home state from viewing their profile. P16 explains \add{that on some sex work platforms, ``you can block certain parts of the world or parts of the country. So I will do that for my state and a state I have family in.''} When such affordances were not directly available, workers manually took care to avoid following anyone connected to their personal account and vice versa. \add{P9 describes taking care to keep ``my personal social media and my sex work social media very separate'' in order to avoid a situation where a social media site ``starts saying maybe you want to follow this person because they're a mutual.'' }
\add{This approach also mirrors "strategic outing"- an approach discussed in prior work on digital dating~\cite{carlson2020love}.}



\subsubsection{\add{Potential legal risk}}
On the other hand, one participant shared their fear of increased legal risk from distributing pornography, clearly highlighting a major benefit of avoiding DTC sales and instead using platform mediation, since platforms age-verify content buyers. P31 explains that in the major U.S. city in which they reside: 
\begin{quote}
``the repercussions for prostitution are lesser than the repercussions for selling pornography to a minor, so I was like if I'm going to screen for in-person sessions I need to screen for selling customs, because I never would want \add{[a charge for selling porn to a minor]} associated with my record ever.
''
\end{quote}

\subsubsection{\add{Theft \& non-consensual sharing of online content}}
The final risk that participants faced was the risk of their content or services being capped or shared. Capping is the practice of screen recording e.g., a live stream / web-camming performance without the performer's consent~\cite{jones2016get}. Whilst various content platforms have some inbuilt technologies to prevent screenshots or recording, in practice it is impossible to prevent recording from a second device. 
Not only is capping theft from the content creator, often those who illicitly steal creators' content then post the content they have stolen elsewhere. While the reasons for theft of sex worker content have not been studied, prior research on non-consensual sharing of other intimate imagery have detailed the reasons behind perpetrators sharing of this content including desire to cause harm, ``for profit, notoriety, social standing, amusement, voyeurism, or no particular reason at all'' ~\cite{ruvalcaba2020nonconsensual}. Participants described their fears and experiences around losing control of their content and finding it on various parts of the internet:


\begin{quote}``Well, the worst is that you have to have in mind that those videos will be leaked, all people around you are going to get them at some point and they will be forever on the internet... Also in face-to-face work I didn't have my intimacy compromised, or if I had troubles they were going to stay there. On online work I know people will talk for days, they can have photos and videos of everything, and I won't be able to ever delete that, they will stay floating around forever.
'' (P13)\end{quote}
This participant explained that they feared these repercussions not only in the present, but for future careers they might want to undertake outside of sex work.

Participants who had not made explicit video content revealed that they had a sense of privacy tied to doing sex work in-person rather than online. P29 describes this very clearly: ``It's harder to distance yourself from porn than it is from escorting, because there's no video evidence of you escorting.'' 

There were three main strategies that participants used to minimise the risk or repercussions of their content being capped or shared. First, participants reported in engaging in identity management, e.g., wearing wigs, obscuring tattoos, or even using digital tools to edit their photos to look less like themselves. \add{For example, P15 explains that they 
``
``do quite a lot of editing on things like my face, not just blurring, also flipping photos and [other] editing...to try to maximise my anonymity as much as possible.''} 
These approaches overlap with strategies participants and other in-person sex workers use such as advertising without showing their faces~\cite{castle2008ordering,barwulor2021disadvantaged}.


Second, participants described a variety of ways in which they leveraged legal mechanisms such as the Digital Millenium Copyright Act in the U.S., which protects the copyrights of a creator over their digital content, to request their content be removed when it was posted without their consent. To use these mechanisms, however, creators must \textit{find} their content and in some cases, the mechanism requires that the creator in fact expose additional information, (i.e., legal name/address), which then may be visible to the person who stole/shared the content~\cite{HowToSen73:online}. To do so, some creators manually searched popular porn sites, while others set up Google alerts for their screennames \add{and, as P22 describes, they send DMCA takedown notices to platforms ``all the time'' to take down the non-consensually uploaded content their Google Alert identified based on their performer name.}

Participants also reported that there were services that would do this labor for them. However, none of our participants used these services as they described them as being too expensive.
\add{For example, P34's cost-benefit analysis suggested that using such services, "would
make sense if I was making what a top girl would be making and I was constantly turning things out but I wasn't.''}

To ensure ease of proving that the content was theirs when they inevitably had to request for it to be taken down from somewhere it was reshared, and in some cases in an effort to dissuade clients from reposting it, many creators watermarked their images, sometimes even with the name and identifying information of the client who had commissioned it. \add{P11 described their approach to reducing resharing of custom videos:}
\begin{quote}
``I will use their name a lot, so if I’m making a video for Peter I’ll be like hi Peter, blah blah blah, oh, Peter, do you like this? and so I think it’s only for him, he won’t be able to turn around and sell it to someone else.'' (P11)
\end{quote}

 
 Engaging in these protections is time consuming and burdensome. As a result, some participants chose to work \add{on particular platforms because they automatically watermark content or even send DMCA takedown notices on creators' behalf.} 



Despite their best efforts to protect their content, participants were frequently unable to get their content taken down: ``it's really easy... [for] things to end up in websites that are based in other countries that I have no access to and can't get my things removed from that won't listen to DMCA takedown[s]'' (P21).

As a result, participants were forced to accept potential loss of control over their content. As P7 succinctly explained, ``I think if you do online sex work you have to accept that your content will be both stolen and resold, unfortunately.'' Some participants even pro-actively outed themselves to their families who had previously not known about them doing sex work, as a preemptive step to protect themselves from potential malicious actors. Participants expressed sadness and resignation about this lack of control and having to take these steps:
\begin{quote}``Yes, I actually had to grieve... letting go of the feeling of security and anonymity. I was like if I start to do camming someday one day my videos will leak out
\add{...}it's just a matter of time and I just decided to be okay with that and to be okay with the worst possibility.'' (P4)
\end{quote}
%
%

\subsection{Reward}

There were three main categories of benefits that participants reported from shifting to online-only sex work at this time: financial (either increased earnings from a wider client base, flexibility/passive income or simply survival when there were few opportunities for in-person work); physical (avoidance of harms like rape, murder or assault and obviously COVID, plus the avoidance of potential interactions with law enforcement in places where in-person sex work/breaking lockdowns was criminalized); and self-expression/branding: either increased freedom for self-expression outside of normative gender or beauty standards or a space in which to improve or expand the sex worker's in-person sex work brand.

\subsubsection{\add{Financial}}
Financial benefits varied widely across participants. Interviewees who described either already having the skills to do online work or enjoying the process of content creation tended to report higher earnings. Some participants earned very little from online work and either drained any savings from previous work, or relied on alternative sources of income such as government benefits, mutual aid funds, or help/loans from family, friends or clients. 
\add{For example, P26 reported only making "a third to at most half'' of what they were making the year prior.}

Beyond specific earnings, participants reported appreciating the potential for online-only work to generate passive income, which was derived from making a piece of content and then selling it many times. Some interviewees reported a vastly increased and/or more consistent income, due to the fact that their client base and earning potential were far greater online than they had been doing face-to-face work. 
\add{P27 described the increase as also being `` a lot more regular, so like a salary every month,'' and that averaged over the year, it was '' 4 times what I made before.'' }

Similar to other workers who transitioned to remote work as a result of the pandemic (see, e.g., \cite{bezerra2020human})
\add{ and those more broadly employed in the offline platform-mediated gig economy~\cite{hunt2019women}}, our participants appreciated the increased flexibility in \textit{how} they earned their income. This flexibility took the form of either passive income, as aforementioned, or flexibility in doing work at home at any time of day convenient to them. \add{For example, P33 explained that 
%
``it was nice to be in my pyjamas and quickly log in to have a meeting online and then quickly log out and [not] have to get on public transport for 30 minutes to get somewhere.''}
\subsubsection{\add{Physical safety}}
Since online sex work generally is done in isolation (with the exception of creating partnered content or employing a camera-person), the physical risks of in-person sex work (sexual assault, rape, murder, etc.~\cite{mcdonald2021s,sanders2018internet, gill2012violence, monto2004female, kinnell2013violence}) are eliminated\add{. As P21 explains, with online work %
 ``I don't have to worry about getting mugged or raped,'' with in person work, ``
 I was always very careful about it but there was always that worry at the back of my mind.''}

This risk was exacerbated during the pandemic, because client numbers dropped drastically~\cite{azam2021covid} as physical lockdowns prevented movement of people either across borders, to their normal workplaces or even from leaving the house. As a result, the clients who requested in-person appointments were lower-quality and posed more physical threat to workers. \add{As one worker explained, with in-person work, especially in criminalized contexts,}
\begin{quote}
``
already people can boundary push because they think we're doing something illegal so we're not going to go to the police and tell on them but the lockdown restrictions added another layer to that where they thought we were also doing something wrong \add{[by] meeting people
...} during the pandemic I had a lot more violent or...unpredictable [clients]...'' (P9)
\end{quote}
Further, even in places where sex work was previously regulated or legalized, sex work was temporarily criminalized during the pandemic either by law or de facto by restriction of movement injunctions, thereby making previous legal work illegal, and any sex workers continuing or trying to do in-person work became vulnerable or visible to police.

Finally, the primary physical risk that participants hoped to avoid in doing online-only sex work during this time was catching coronavirus. This was especially critical as sex workers are especially likely to be disabled~\cite{fritsch2016disability} and, accordingly, 50\% of our population self-described as disabled. 
\add{P15 had to take extra precautions during the pandemic to protect their household and described online work as a "life saver."
}

%
%
%
\subsubsection{\add{Self expression \& branding}}
Participants who felt pressure to conform to beauty and/or gender normative standards described doing online-only sex work as having greater possibilities for experimenting with different looks, changing their look to be more inline with their personal gender or beauty goals or being free to express their inner selves in a much larger, more international marketplace that has a customer for every type of sex worker. The global nature of the platform allowed our participants to reach clients they would not otherwise have had access to. \add{P7 explains that online sex work:
\begin{quote}
 ``
 helped me feel more comfortable about my body, especially as a trans woman going through my transition...
 even if they fetishise you they want to see your body and [that] can be quite affirming.''
 \end{quote}}

Some participants described that they felt the period of continuously or at least regularly making photos or videos of themselves improved the branding for their in-person work. It can be difficult to find the time or energy for good quality selfies or photoshoots when you are travelling or doing high energy in-person client work. 
\add{P2 described this effect, saying "I have to keep producing content to keep the fans happy, and actually that is better for my escort business"; adding that they could even gain a third usage for the material by directly sending it as special bonuses to "[high paying] clients... who expect a really personal relationship."} 

\subsection{Future Plans}
In general, the success of their online-only sex work in terms of enjoyment but particularly income correlated with participants' expressed interest in doing online sex work in the future. Many participants expressed that they saw online-only sex work to be a useful and productive adjunct to their future in-person work. Some saw online work as a way to enhance their existing in-person work: either deepening their branding or providing a way for potential or regular clients to connect with them before or between in-person sessions. 
\add{P29 described the integration of their online and in-person work, saying that in-person ``clients who had already seen me would interact with me online and then prospective clients that were maybe a bit cautious or not sure would also connect with me online.''}

Others saw online work as a way to reach a wider client base: clients in differing geographic regions or with different levels of disposable income limits (clients who could not afford an in-person session might still contribute to a sex worker's income by buying content or a subscription). \add{P26 spoke of online sex work in terms of remote work, calling it "the future of sex work", and as a business analysis said ``I really see the value in just having content for sale that accompanies all the other work you're doing.''} 

However some participants described that online work had affected their branding in a negative way. Whilst not all content creation in this space is pornographic, much of it tends to be erotic or explicit. This proximity to a different, more public sector of the sex industry than the one that participants had originally chosen had real or imagined repercussions in their client relations. 
\add{P2 felt concerned about potential clients being "put off by what they see" but neatly summarized by saying "I can't pay my bills on the basis of what those people think so..." }

In sum, 20 participants planned to return to in-person work but also continue online-only work, 4 were unsure about their future plans, 6 planned to stop online work and return to strictly in-person work, and 3 participants who had previously done mostly in-person work had little intention of returning to it at all. 
\add{P13 explained ``I get more money from online than in person ever, so I don't really feel the need to do it in person.''} 

Finally, one participant, P3, chose this moment to leave the industry entirely: ``Officially I have retired, actually. So, that's a big change to my in-person work.''

\section{Discussion}
Qualitative interviews with 34 participants who had previously done in-person sex work and pivoted to online-only sex work as a result of the Covid-19 pandemic revealed a number of challenges and benefits associated with online-only work. In the face of the natural experiment forced upon sex workers by the pandemic, it is tempting to ask whether, based on the data collected, we could say that online work was ``better'' than in-person work. It is impossible to quantify whether sex workers faced ``less'' risk by doing online work since the baseline ``risk'' involved in sex work is both under-researched and highly specific to each type of sex worker, the specific context in which they are working and many other factors~\cite{sanders2018internet}.
While we cannot say that online work involves greater or fewer risks than doing in-person work, we find that our participants reported different, new and increased online risks as well as workplace benefits as a direct result of their switch to online-only from in-person work. 

Participants' primary challenges were the risk of account deletion/platform loss and loss of control over the content they were sharing online. The primary benefits participants reported were reduced risk of physical harm (including Covid) and financial welfare, with some participants earning more than they did with in-person work and others earning the same amount or less, but enough to sustain themselves.

The most interesting indicator about the rewards of online sex work is to be found in considering that a large proportion (23 participants) planned to continue with digital work when they could conceivably return to in-person work, either to diversify their income possibilities, or as a wholesale switch. Some participants planned to continue due to the sunk privacy cost of switching to online work during the pandemic: since they had already revealed information about themselves that they perceived it would be hard to erase (i.e., creating and distributing explicit material containing their face) due to needing to work online as a result of the pandemic, so they felt they might as well continue to profit from it. Others were already comfortable with the public sharing of explicit content required for online work because they had previously made studio porn. The former illustrates how privacy fatalism~\cite{xie2019revealing} and lack of control over digital content intersect with the stigma of sex work~\cite{benoit2018prostitution} to reduce worker autonomy. In Section~\ref{discuss:digitalimp} we discuss further implications of lack of content control.

\add{Results of this research can be used by sex workers to upskill and improve their business practices. Not all sex workers will be aware of all the strategies discussed here and whilst in this paper anonymizes platforms and tools, community members will be able to seek out or ask colleagues for more information once they are aware of what kinds of tools and strategies available or possible. Sex worker support services may use this research as part of resources and strategy guides designed to help sex workers navigate online work as safely as possible.\footnote{See for example: \url{https://hackinghustling.org/resources/}.} Further, sex worker rights groups can use our work to discuss the risks and benefits of digital sex work, and ways to mitigate those risks, with policy makers and platform developers as part of efforts to make a safer working experience for online sex workers.}

Finally, our discussion would be incomplete without mentioning the pride many participants felt in their resilience to the additional challenge of transitioning to an entirely new modality of work in their already stigmatized and marginalized profession and not only sustaining themselves but thriving:  
\begin{quote}
    ``Even though it was difficult, in the pandemic it was positive to know that it's something that I can be flexible with and adapt to. At the end of the day I can adapt and learn a new skill and hustle and survive.'' (P26)
\end{quote} 
%

\subsection{Implications for Digital Experience and Design}
\label{discuss:digitalimp}
Below we summarize concrete implications from our work for the design of safer digital experiences that can enhance the well-being of informal labor market workers, particularly the most marginalized. Additionally, we draw connections between our findings and broader on-going conversations around platform governance, content moderation, and safety. \add{While we suggest opportunities for integration and creation of new platform features, we note that beneficial features already exist in some sex work platforms. Integrating these existing features throughout the digital sex work ecosystem is a valuable first step, in addition to the design directions we suggest. The platform features that are already helpful for sex workers are: technology that prevents screen recording; content watermarking;  the capacity to block a state/country of the sex worker's choosing; and easy blocking for bad or abusive clients.
}

\subsubsection{\add{Harassment}}
Our participants reported frequent interactions with clients who harassed them. In-person sex workers typically screen new clients using a variety of screening methods~\cite{barwulor2021disadvantaged}
\add{as do those engaging in meeting people offline from dating apps~\cite{albury2019safety}.} However, the current platform technologies available to online sex workers do not allow pro-active screening. Instead, a sex worker must encounter harassment -- with the individual risk that the client then takes the harassment offline -- before being able to protect themselves.

Some advertising directories where in-person sex workers arrange appointments have integral client ``vetting'' systems, where a sex worker can leave a score or comment on the client after an appointment, which can help a subsequent colleague make an informed choice as to whether they wish to accept the client or not. Additionally, some in-person workers will require a good reference from an established colleague that the client has previously seen, as well as or instead of ``real-world'' information such as a work email address or professional social media link~\cite{barwulor2021disadvantaged}.
In this context, online-only sex work platforms' current approach to client behavior management places the broader online sex working community at risk, since an individual worker blocking an individual client only protects that individual sex worker. Similar issues face other workers engaged in platform-mediated gig work, as documented in prior work on rideshare drivers~\cite{almoqbel2019individual}
\add{and app-based beauty workers~\cite{raval2019making}. In sum, we can see many different kinds of workers (but especially gender-marginalised people) engaging in what Gillett called "safety work", in a study of online dating~\cite{gillett2021not}. Security practices connect and overlap between commercial and non-commercial types of engaging with strangers.} 


In addition to a lack of proactive screening as a potential reason for the greater harassment our participants reported during online work, we hypothesize that clients may have a more depersonalized relationship with online-only sex workers than those they see in person. \add{Prior work consistently suggests that sexual harassment online is overwhelmingly directed at women~\cite{henry2018technology}, who comprise a majority of the sex working population and our sample. Further, }
prior work on harassment in other spaces suggests that depersonalising others contributes to harassment~\cite{fox2017women,postmes2002intergroup}.
Men may be especially likely to depersonalize those of other gender identities; men make up the majority
of sex work clientele~\cite{monto2004female}. Finally, as is noted in many studies of online harassment, including those specifically focused on harassment by those who identify as men, anonymity greatly increases harassment~\cite{rubin2019fragilemasculinity,blackwell2018online}. In digital sex work settings clients are typically anonymous.
%

Power over online speech and harms is increasingly concentrated in the hands of platforms, especially in the adult industry~\cite{tiidenberg2021sex}. There are a number of potential solutions that platforms could leverage to proactively protect workers from harassment, many of which are currently implemented on other types of platforms or have been proposed in non-labor contexts. As Schoenebeck et al. note, reactive, carceral approaches to harassment management are pervasive across digital platforms~\cite{schoenebeck2021drawing}. Their participants (who were not explicitly digital workers nor sex workers) favored strategies such as apologies from harassers and, in line with sex workers' existing ``bad client lists,'' offender lists. Drawing on the latter, platforms that mediate digital labor, such as sex work, could consider digital fingerprint-based blocking (e.g., IP banning) of harassers from particular platforms. Of course, the latter does not stop a bad client from utilizing circumvention tactics.

Additionally, Tinder proactively warns a user sharing their phone number that this behaviour might entail certain risks~\cite{website:axios}; Twitter now proactively signposts that certain conversations can get heated and proposes caution in inflammatory posting, in light of the fact that most users will be real people~\cite{website:metro}.
Digital labor platforms could similarly act on poor client behaviour, with suggestions or prompts that the sex worker is a real person and should be contacted with respect: a warning like ``did you really want to send this one word message?'', ``please don't put the creator you're contacting at risk by mentioning acts or services illegal in your jurisdiction,'' ``remember that the creator you are contacting is a real person, keep your conversation delightful and polite!'' Platforms could also draw on the existing strategies of in-person sex workers and platforms such as Clubhouse~\cite{HowtoGet14:online} to use a reference-based system for allowing new clients to join. Alternately, platforms could offer graduated access to platform features that can create greater threat (e.g., the ability to send photographs or launch a video call). We note that, as discussed further below, such solutions should be developed and evaluated in collaboration with all stakeholders; in the context of our study: online sex workers and their clients. 
%

\subsubsection{\add{Burnout}}
 Many of our participants reported difficulties with work-life balance and burnout as a result of switching to online work. Digital platforms can support worker well-being by incorporating carefully designed nudges and opportunities for their creators to scale their efforts. For example, evolving tools for passive income. For digital sex work platforms, such tools could include an easy way to create clips from live shows and creating resources to educate sex workers on scheduling posts or setting boundaries with clients to avoid the feeling that they need to be online constantly: ``people were expecting instant replies all the time from me, they would write me at any hour, out of office hours and then when I don't reply they send me two more emails'' (P30). 
 
 Platforms also have the opportunity to offer a venue for worker peer support, something that is difficult for digital gig workers to access due to lack of formal workplace environment~\cite{rivera2021want} and for sex workers to access because many digital platforms prohibit the discussion of commercial sex work~\cite{barakat2021community}. 
\add{Delivery drivers~\cite{seetharaman2021delivery}, rideshare drivers~\cite{almoqbel2019individual}, beauty workers~\cite{raval2019making}
and domestic workers~\cite{hunt2019women} much like our participants, use private messaging groups for watercooler chat as well as more serious functions like helping keep each other safe. We propose in-app messaging between workers, forums or other platform-sponsored communication affordances so that workers aren't isolated. We posit, in line with Barakat and Redmiles~\cite{barakat2021community} that workers want to communicate with each other for the purpose of maximising profit and minimising harm. Both of these together contribute to minimising burnout, which occurs across the gig economy. Thus,} by creating a venue for their workers to network and share tips, platforms can improve both worker well-being and productivity, which is mutually beneficial.
 
Finally, some digital platforms now prevent extended usage: Tinder and Bumble limit the amount of free swipes per day; Tiktok auto-plays an anti-``doom-scroll'' message after a specific amount of time spent on the app. Whilst these examples are not (generally) workplaces and therefore a similar intervention would need to be carefully managed to avoid loss of income, digital wellness or hygiene suggestions could be made available to creators in light of the very high work burden and common burnout that they are reporting.



\subsubsection{\add{Intellectual property}}
A significant challenge that faced our participants was lack of control over the content (typically, sexual images or videos) that they created and sold. This challenge faces others who share intimate images within the context of a relationship (i.e., those engaged in non-commercial sexting) as well as those who create paid non-sexual content such as digital artists, musicians, and movie creators. The music and film industries go to great lengths to protect their intellectual property and platforms have entire teams dedicated to taking down such copyrighted content~\cite{horwitz2021facebook}. 
Additionally, platforms such as Facebook have attempted various iterations of software to address the sharing of non-consensual intimate imagery (NCII), which is typically sexual images or videos that someone shared with a specific individual and which have come to be posted publicly online~\cite{Metaista61:online}. While early iterations of an approach to deal with this received understandable pushback as they required the victim to submit their content to a third-party database, increasing their privacy violation, a new approach allows victims to fingerprint their content without it leaving their device. Those fingerprints are loaded into a database against which digital platforms can query to ensure that content being uploaded on their site is not NCII.

%

Just because sexual content is created for commercial purposes does not mean the creator has consented for that content to be shared beyond their original intent. Such non-consensual sharing is not only theft, it also creates significant privacy violation, particularly due to the intimate nature of the content. Further, given the stigma of sex work~\cite{benoit2018prostitution}, the ability to maintain control over content is necessary for workers' autonomy: their ability to transition to a non-sex-industry job, to obtain housing, and in some cases to avoid violence~\cite{mcdonald2021s}. 

Material on advertising directories and social media accounts maintained by in-person sex workers tends to stay in their control and can be removed when the sex worker leaves the industry, which is a sharp contrast to the potential complete loss of control participants reported feeling over the explicit content they produced for online-only sex work. In addition, the ability to control their own materials may allow sex workers to invoke data rights over its removal from the internet, such as the GDPR Article 17 `right to be forgotten'~\cite{Righttoe99:online}. 

Sex workers' existing approaches to content ownership include watermarking content  and using reverse image search or services that use similar techniques to find their non-consensually shared content and request it be taken down. Technological innovation around cryptographic watermarking (ideally hardened to prevent removal of watermarks even from screenshot content) could improve the efficacy of the former approach and privacy-preserving fingerprinting approaches similar to those explained for NCII above could improve the efficacy of the latter.

\subsubsection{\add{ Deplatforming and transparency}}
The enormous number of strategies sex workers employ in attempts to avoid platform loss speaks both to workers' intense fear of deplatforming and the severity of its potential repercussions. Ours is not the first study to document the existence and consequences of deplatforming, both in the sex work community~\cite{blunt2020posting, are2021shadowban, are2021sex} and gig work more broadly~\cite{website:cnbc.com}. Participants recognised that moderation was a complex issue and all they asked for was transparency and sensitivity. This requires a much greater investment in good quality moderation - something which even non-adult platforms currently struggle with - but which must be implemented if platforms are not to continue to marginalize the vulnerable. This issue in online sex work exactly mirrors the problems discussed in social media content moderation by Díaz and Hecht-Felella~\cite{website:brennancenter}, Haimson et al.~\cite{haimson2021disproportionate}, Schoenbeck et al.~\cite{schoenebeck2021drawing} and many other scholars.

Our work, to our knowledge, uniquely documents the measures that workers -- particularly those who must advertise their services across platforms -- are taking to avoid deplatforming. Specifically, our participants reported using algorithmic transparency tools, an example of which is \url{nsfwjs.com}, a tool that identifies portions of a photograph that an automated content moderation algorithm may consider too explicit and flag for removal. Such tools allow workers to avoid account ``bans,'' which escalate in length from 24 hours to permanent account removal based on the number of prior offenses a user has had in a given time period~\cite{seering2020reconsidering}. Even a 24-hour ban from e.g., posting content designed to advertise a sex worker's digital services can have severe consequences for those in financially precarious situations. While external (e.g., researcher-created) algorithmic transparency tools are a promising harm-reduction approach for future work, platforms may also consider offering such tools themselves. Arguably, users who wish to avoid deplatforming (or temporary account bans) and platforms who seek to ensure content follows their terms and conditions are unified in their goals. Thus, to more effectively meet the goals and needs of both platforms and creators, we suggest the need for a shift from the existing carceral approach to content moderation -- where infringements are punished after the content is posted and subsequently removed -- toward a cooperative and transparent approach by offering platform-created content-checking tools that creators can use to avoid post-hoc algorithmic moderation
. Notably, however, such tools will not address report-based and human moderation, for which there exist numerous transparency proposals~\cite{de2020democratising,jhaver2019does,suzor2019we}. 

Relatedly, participants reported that it was incredibly complicated for them to keep up with the ever-changing, often vague, and largely unexplained regulation changes on digital sex work platforms: what was and was not permitted in terms of content on any particular platform. 
Whilst Facebook's X-check program garnered criticism for its elitist favouritism of celebrities and conservative politicians~\cite{website:TheWallStreetJournal}, a considered program of hands-on account management allotted to creators who might be more at risk of skirting close to moderation regulations might be a sensible addition, particularly for platforms through which creators earn a living and from whom platforms particularly profit. This could take the form \add{of a clinical approach such as that adopted in prior work with IPV victims~\cite{havron2019clinical} and/or the use} of specialised liaison managers such as the account managers that are allocated to high profile Instagram creators (one of our participants had such an account manager). 

Finally, as noted in prior work on content moderation, context and subject-matter expertise on the part of content moderators and algorithm designers (e.g., fluency in a particular language, appropriate cultural competence and context) are critical for fair moderation~\cite{suzor2019we}. Similarly, digital sex work platforms must consider whether their algorithm designers and content moderators are well-versed in the consent norms, range of sex acts and practices, and many genders and sexualities of the creators and users they seek to moderate~\cite{stardust2018safe, paasonen2019nsfw}. To ensure such expertise and fair moderation, platforms may seek to employ expert community insiders to partner on design and moderation efforts, as discussed further in \add{Section 5.3}. 
%
\add{\subsection{The Informal Labour Landscape: Digitisation}}

\add{Workers in the informal labour economy are increasingly turning to platform-mediated work across the world. Those engaged in often informally structured in-person labor such as taxi drivers become Uber\footnote{https://www.uber.com/} drivers, beauty workers become Stylebee-s\footnote{https://www.stylebee.com}, artisans become Etsy-ers\footnote{https://www.etsy.com} and domestic workers become SweepSouth-ers\footnote{https://sweepsouth.com} and Hilfrs\footnote{https://hilfr.dk/en}. 
Others in the growing informal labour economy have converted their existing desk jobs to platform-mediated freelance opportunities through platforms like UpWork\footnote{\url{https://www.upwork.com/}}. Yet other informal labourers are working at newly created digital labour such as crowdwork~\cite{gray2019ghost}.} 

\add{Our participants intersect with yet are unique from each of these groups of digitally-mediated informal labourers. 
Our sex-working participants are taking their previously in-person labour and moving it entirely to a digital workplace, where both the labour and its mediation is conducted digitally, a feat impossible in many other kinds of in-person informal labour, which have become platform mediated but not fully digital. Moving to platform-mediated work for other informal labourers has the same kinds of motivations as it does in some places for our participants - increased flexibility, increased income and greater access to clients that they might not already have~\cite{hunt2019women, raval2019making, seetharaman2021delivery}. 
However, this comparison more accurately reflects what happens when in-person sex workers who have previously been working in venues or for managers move to online advertising~\cite{sanders2013sex}}

\add{Online-only sex workers are in many ways more similar to those engaged in feminized digital freelance work (e.g., graphic design, journalism) in the nature of their labor -- advertising strategies, building of their own personal brands, creativity in producing primarily digital products and services -- as well as the economic precarity and both emotional and affective labour inherent in their work~\cite{salamon2016lancer,makinen2021resilience}. While women are overrepresented in the informal labour market generally~\cite{elgin2021informality,beneria2001shifting}, digital sex workers experience unique stigma due to the sexual nature of their work and, intersectionally, the prevalence of women, disabled and queer people in the sex-working community.} 

\add{The digitization of both sex work and freelance work traditionally confined to formal workplaces has democratized this labour~\cite{ferrucci2020boundaries}. As one of our participants noted, digital sex work services are significantly cheaper than in-person services and, as such, a far wider range of clients can afford services. The flip side of this democratization for all forms of labour, sex work or otherwise, is often reduced wages and eroded worker rights~\cite{makinen2021resilience, salamon2016lancer}, as we discuss next.}\\
%
%
%


\subsubsection{\add{Regulation and workers' rights}}
\add{We note that while technology can both create and prevent harm, so too can legislation. The potential for technological intervention would ideally be accompanied by social and legal changes, across both the space of digitally-mediated informal labor and the digital intimacy space~\cite{freed2019my}.} 

\add{There is ongoing debate regarding the employment status of digital gig workers. Prior work outlines various proposals for regulating gig work, many of which center on reimagining the definition of employees, employers, or work itself~\cite{stewart2017regulating}. As a way toward this reimagining, Van Doorn proposes ethnography and platform cooperativism as potential future directions for centering gig workers in improving working conditions~\cite{van2017platform}. If online-only sex workers are employees (as prior work suggests they may be~\cite{marston2020onlyemployees}), then the issue of account deletion and/or platform loss becomes a workers' rights issue, irrespective of all the other potential benefits/losses employee status might bring to content creators~\cite{berg2020porn}.} 

\add{While the digitally-mediated platform space remains largely without strong worker rights protections, some changes are beginning to occur. For example, in Denmark, one platform for cleaners signed an agreement for worker rights such as sick pay and pensions~\cite{ilsoe2020hilfr}. Similar initiatives could be implemented across other platforms, since they might offer more flexibility than legal solutions. However, movement toward worker rights may create harm in the near term. As Hunt et al. concisely explain that one of the key issues in platforms failing to provide worker rights is perception of themselves as "technology platforms, not employers"~\cite{hunt2019women}. Thus, when regulations of digitally-mediated work are instituted platforms may pull out of the market leaving workers scrambling for new labour opportunities.\footnote{\add{See for example Spain's attempted regulation of workers rights for delivery drivers and resulting impacts~\cite{website:huckmag.com}}}}

\add{Finally, complicating this already nuanced landscape of gig work regulation is the sexual nature of the work our participants undertake. Existing laws regarding distribution, production, and sale of sexual content do not address the realities of digital sex work nor appropriately protect these workers~\cite{stardust2018safe}. Further, appropriate legal protections are lacking for all forms of digital intimacy with victim blaming and stigma perpetuated by justice systems being pervasive in situations of harm within digitally intimate contexts whether for work or recreation~\cite{lageson2019gendered}.}\\
\subsection{\add{Participatory Action}}
Throughout this discussion we have noted many opportunities for potential technological interventions to improve the well-being of online-only sex workers, as well as those engaged in digitally mediated-gig work and who make a living from content-creation in general, regardless of whether that content is sexual in nature. We note, however, that a participatory action or design justice approach should be taken for any instantiation of these proposals or other interventions~\cite{costanza2020design}. As digital workplaces, digital sex work platforms may consider how to scale such participatory action, for example through user feedback groups, interfacing with elected community representatives, etc. 

\add{In the development of both policy and technology i}t is critical that the voices of online-only sex workers are included with the voices of other digital gig-workers to ensure the well-being of all workers.
To ensure this inclusion will require a shift from existing adversarial norms toward digital sexual content creators~\cite{sas_paper, website:medium.com} toward respecting these creators as digital workers and stakeholders of the digital spaces in which they conduct their work.
%
%
%

\section{Conclusion}
 \add{In this work we interviewed 34 sex workers about their experiences pivoting from providing primarily in-person direct-to-client services to online-only work as a result of the COVID-19 pandemic. Online, our participants found new benefits -- increased physical safety, flexibility, and in some cases freedom of self-expression -- and also new risks: greater digital exposure, burnout, and loss of control. Our findings suggest several opportunities for improved platform design to better protect online-only sex workers and others engaged in digital intimacies.} 
 
 \add{Online-only sex workers sit at the intersection between growing conversations regarding safety work, content moderation, particularly of sexual content, and digitally-mediated labor. While they share experiences with other groups of digital labourers, online-only sex workers are unique in the multiple roles they may simultaneously find themselves in (e.g., porn producer, worker, employer, social media marketer) and the stigma they face in their work.  We underscore the criticality of including sex worker voices in these ongoing conversations, particularly in light of the cross-cutting nature of their experiences across digital intimacy and digital labour.}\\


\section*{Acknowledgements}
This work would not have been possible without the grace, time and labour of the sex workers that took part in the research: to everyone we spoke to: thank you so much for your trust. Also the transcriber P, who is rad. 

The second author completed a portion of this work while at the Max Planck Institute for Software Systems.

\bibliographystyle{acm}
\bibliography{unblinded}

\clearpage
\appendix


\section*{Supplementary material}

# A  Interview Protocol

The interview protocol of questions analyzed in this work is presented below.

**PRIOR TO COVID**
- Tell me a little about your background? What type of work did/do you do?
- If you're comfortable sharing, how long have you been doing sex work?

**WEB-MEDIATED WORK IN GENERAL**
- Are you currently offering online services? What services are you currently offering?
- How did you decide what online services to offer?
- How did you decide how to price the service/s you are offering?
- Has any of the services/prices changed since the beginning of the pandemic ? If so, how?
- How would you describe your clients? [Prompts if needed: Are you selling to previous in-person clients, people completely new to you, people who you hope to convert to in-person clients post-pandemic, or a mix?]
- Do you perceive that offering online services has affected the persona you use for in person work (your "brand")? If yes, how will you manage this /are you managing this?

**PLATFORM WORK**
- Are you using one or more platforms to sell your content? If so, which one/s?
  - How did you choose the platform/s you use? Were there any platforms you considered but decided not to use? Or platforms you tried and stopped using? Why?
- Do you do anything to try to optimise how much your content is seen? [prompt algorithm, schedule, engagement groups, multiple posting, hashtags]
  - If yes, please describe
- Do you do anything to avoid censorship or your content being hidden? [prompt blurring/censoring body parts, censoring keywords]
  - If yes, please describe
- Has anything changed about the platform while you have been using it? What changed? How has this affected you?
  - [If not volunteered, AND relevant] Has anything related to being paid changed on this platform during this period? Have the payment or legal conditions of the platform changed during this period? If yes, how has this affected you?
  - [If not volunteered] Has the legal situation of the platform changed during this period? How has this affected you?
- Have you read the terms of service of the platform(s) you're using?
  - If no, do you think that the work you're doing is allowed under the platform's terms of service ?
  - If yes, is the work that you're doing allowed under the platform's terms of service ?
- If the work is not allowed: what do you think are the risks of doing this work?
  - Have you taken any actions to avoid these risks? What actions?
  - If the work that you do is against the terms of service of your platform (eg camming through Skype, sharing porn through google drive), how do you protect yourself from losing access to the platform?
- If work allowed, how would it affect you if the platform changed its terms of service to exclude your type of work?
- If you were no longer able to use this platform, what would you do instead?
- How do you advertise your work?
  - If using social media:
    - Have you read the terms of service of the platform(s) you're using?

- - If no, do you think that the work you're doing is allowed under the platform's terms of service?
  - If yes, is the work that you're doing allowed under the platform's terms of service?
  - If the work is not allowed: what do you think are the risks of doing this work?
    - Have you taken any actions to avoid these risks? What actions?
  - Have you noticed any changes in what is allowed or not on your advertising outlets? If yes, has this affected your work? If yes, how?

**ABOUT THE WORK**
- Have there been any benefits, expected or unexpected, in moving to online work?
- What are the challenges/issues you faced in moving to online work?
- Are you still facing those challenges/problems?
- Have you used any automation in your work?
  - Prompts if needed: chatbots, autoresponses or anything where the tech does some part of the work for you?
  - Why do you use/not use automation?
- Are there any resources or tools that do not exist, but you feel would make your work easier?
- Have you felt "burnt-out" from work at any point during lockdown[ since March 2020] ?
  - Had you experienced this feeling when you were conducting work before lockdown[doing sex work before the pandemic] ?
    - If yes, was the burn out you've experienced recently different? In what ways?
  - Did you change your behavior or how you were doing your work as a result of feeling burnt out?

**INCOME & PAYMENT**
- How has your total income changed due to the pandemic/moving online?
- Have you received financial support (eg from government / mutual aid/family/ friends/ loan from a client etc) during the pandemic? has this affected how/how much you work online/ not?

**SECURITY/PRIVACY**

- Have your concerns about safety changed with your switch to online work?
- If you sell content (as opposed to only live work), how do you protect your content from being re-sold by the buyer?
- If you cam/stream, do you use anything to prevent your work being recorded/shared?
- Do you worry about people who don't know you do sex work finding out from your online work? How does this differ from before?
- Do you take any steps to avoid outing? What are they?

**THE FUTURE**
- Do you intend to keep offering online services once it is possible for you to return to in-person SW?
- What are the factors that will make it possible for you to return to in-person work? Do you have any anticipation of when this might be?
- Overall how has your experience been of online work?
- Is there anything else you'd like to tell me about your experience of pivoting to online work during the Covid-19 crisis?

**DEMOGRAPHIC QUESTIONS**

What is the age of your persona (can be age group eg 25-30)?
- Is this the same as your actual age? If no, what is your actual age?

What gender do you work as?
- Is this the same as your actual gender? If no, what is your actual gender?

What is your perceived ethnicity or the ethnicity that you work as?
- Is this the same as your actual ethnicity? If no, what is your actual ethnicity?

What did one hour of the work that you did prior to the pandemic cost to the client?

What country do you mostly work in?

Are you a migrant?

Do you identify as LGBTQIA at work? personally? both? neither?
Do you identify as disabled? at work? personally? both? neither?



\end{document}